\documentclass[aps,pre,twocolumn,superscriptaddress]{revtex4-1}  % for review and submission

\usepackage{graphicx}  % needed for figures
\usepackage{dcolumn}   % needed for some tables
\usepackage{bm}        % for math
\usepackage{bbold}
\usepackage{amssymb, amsmath}   % for math
\usepackage[usenames, dvipsnames]{color}
\usepackage{float}

\begin{document}
\title{Taking snapshots of a quantum thermalization process: emergent classicality in quantum jump trajectories}
\author{Charlie Nation}%
\email{C.Nation@ucl.ac.uk}
\affiliation{Department of Physics and Astronomy, University College London, London WC1E 6BT, United Kingdom}
\affiliation{%
Department of Physics and Astronomy, University of Sussex, Brighton, BN1 9QH, United Kingdom.
}%    
\author{Diego Porras}%
 \email{D.Porras@iff.csic.es}
\affiliation{%
 Institute of Fundamental Physics, IFF-CSIC, Calle Serrano 113b, 28006 Madrid, Spain 
} 
\date{\today}

\begin{abstract}
We investigate theoretically the emergence of classical statistical physics in a finite quantum system that is either totally isolated or otherwise subjected to a quantum measurement process. We show via a random matrix theory approach to nonintegrable quantum systems that the set of outcomes of the measurement of a macroscopic observable evolve in time like stochastic variables, whose variance satisfies the celebrated Einstein relation for Brownian diffusion. Our results show how to extend the framework of eigenstate thermalization to the prediction of properties of quantum measurements on an otherwise closed quantum system. We show numerically the validity of the random matrix approach in quantum chain models.
\end{abstract}

\maketitle

\section{Introduction}

The emergence of an effective classical statistical description of the dynamics of a closed quantum system is an important open question at the heart of the foundations of statistical physics \cite{Bartsch2011, Ates2012, Tikhonenkov2013, Niemeyer2013, Niemeyer2014, Gemmer2014, Schmidtke2016}. The study of quantum non-equilibrium dynamics only recently become experimentally feasible \cite{Schreiber2015,Clos2016a,Kaufman2016, Neill2016, Neill2018, Kim2018}, raising questions surrounding the process and conditions in which isolated many-body quantum systems equilibrate to a thermal state \cite{Gemmer2001, Reimann2008, Reimann2010, Deutsch2010, Short2011, Ikeda2011, Marek2016, Farrelly2017, Borgonovi2017} - a process known as quantum thermalization \cite{Rigol2008, DAlessio2016, Gogolin2016, Mori2018, Deutsch2018}. 
Important related questions remain surrounding relaxation time-scales and the route to equilibrium of complex quantum systems \cite{Garcia-Pintos2017, Richter2018, Schiulaz2019, Alhambra2019, Nation2019, Dabelow2020, Borgonovi2019, Nickelsen2019a, Nickelsen2019}, 
as well as the emergence of thermodynamical laws \cite{Hinrichsen2011, Ziener2015, Bisker2017, Manzano2019}. 
A useful approach to the description of generic non-integrable quantum systems can be developed from quantum chaos \cite{Berry1977, Srednicki1994} and the eigenstate thermalization hypothesis (ETH), which in turn
can be derived from an underlying random matrix theory (RMT) \cite{Nation2018, Nation2019, Deutsch1991, Reimann2015, Reimann2015, Ithier2018, Dabelow2020}.

Most works on quantum thermalization dynamics focus on the evolution of expectation values of local operators, $\langle O (t) \rangle$. How and when the unitary evolution of such observables can be shown to be described by an effective classical Markov process is an important question in the foundations of statistical physics. The main result of this work is to analytically derive such an effective classical theory for quantum equilibration dynamics. Concretely, we show that under physically reasonable conditions the unitary quantum dynamics of a system initialised in a pure state may be shown to be described by an effective Brownian process at a finite temperature. 
Additionally, we analyse a more typical experimental protocol: a set of quantum measurements at times $t_1$, $t_2$, $\dots$, generating a set of outcomes $O_1$, $O_2, \dots$.
Here, a few natural questions arise: Do the observation outcomes have the properties of a classical stochastic trajectory in the appropriate limit?  How do thermodynamical properties of stochastic trajectories emerge within the RMT and ETH picture? How different are the dynamics of expectation values, $\langle O (t) \rangle$, compared to the set of measurements obtained under continuous monitoring? Answering those questions is not only of fundamental interest, but can also lead to novel ways of characterizing quantum devices.

We address the questions above within the theoretical framework of RMT and quantum chaotic wavefunctions \cite{Nation2018}. 
Firstly, we take the more conventional point of view in which the system evolves up to a certain time, $t$, at which a
quantum measurement of a local operator is performed. 
We focus on the variance of measured values of $O$ after a series of experiments, $\sigma_O^2$, and we show that, at long times, it satisfies the celebrated Einstein relation, $\sigma_O^2 \propto k_B T$, with $T$ the microcanonical temperature, provided certain conditions are met by $O$. This demonstrates an effective description of the dynamics of an observable in a closed quantum system by a classical Ornstein-Uhlenbeck (OU) process.

An additional important result is that the Einstein relation is not only observed for the long-tome observable variance, but also for the observable variance of a single eigenstate, which we label the `eigenstate equipartition theorem'. This result establishes analytically a finite temperature description of individual eigenstates of closed quantum systems. This result provides a link between the ETH picture (naively: `eigenstates behave as thermal states for realistic observables') of thermalization to the emergence of \emph{classical} statistical physics - not only are eigenstate expectation values thermal expectation values (ETH), but eigenstate fluctuations fulfil classical thermal fluctuation theorems.

We then move of continuous monitoring of a local observable during the thermalization process, yielding a set of measurement outcomes, $O_j$, at times $t_j = j \Delta t$. From this description we observe the independence of average dynamics on such measurements, and demonstrate that individual quantum jump trajectories \cite{Garrahan2010} may be described by instances of a classical Markov process.
We also show that there exists a quantum Zeno regime for very short $\Delta t$, in which equilibration slows down.
Finally, our approach can be exploited to measure the density of states (DOS) of the system as the ratio between time-integrated fluctuations of 
$\langle O(t) \rangle$ and the variance of quantum measurement outcomes. 
We numerically check our results in quantum chain models. 

This article is arranged as follows. In section II we set up the scenario under study. This is followed by section III where we introduced the methodology by which we treat non-integrable quantum systems - quantum chaotic wavefunctions, and their description in terms of RMT. In section IV we discuss the emergence of the Einstein relation from chaotic wavefunctions, and in section V we extend the discussion to the case of sequential projective measurements, obtaining a statistical description of quantum jump trajectories of closed quantum systems. In section VI we show some numerical verifications of the theory, before concluding in Section VII. Additional numerics, and proofs, are given in the Appendix.

\section{Set up}
Consider an isolated finite quantum system separated into a `(sub)system' $S$, and `bath' $B$. The system Hilbert space is defined as the support of a local observable of interest. The interacting Hamiltonian is $H = H_0 + V$, with 
$H_0 = H_S \otimes \mathbb{1}_B + \mathbb{1}_S \otimes H_B$, where $\mathbb{1}_{S(B)}$ is the identity on the system (bath) Hilbert space. 
Note that the system Hilbert space can correspond to local degrees of freedom in an homogeneous system, or a system weakly coupled to a finite bath.

\begin{figure}
	\includegraphics[width=0.45\textwidth]{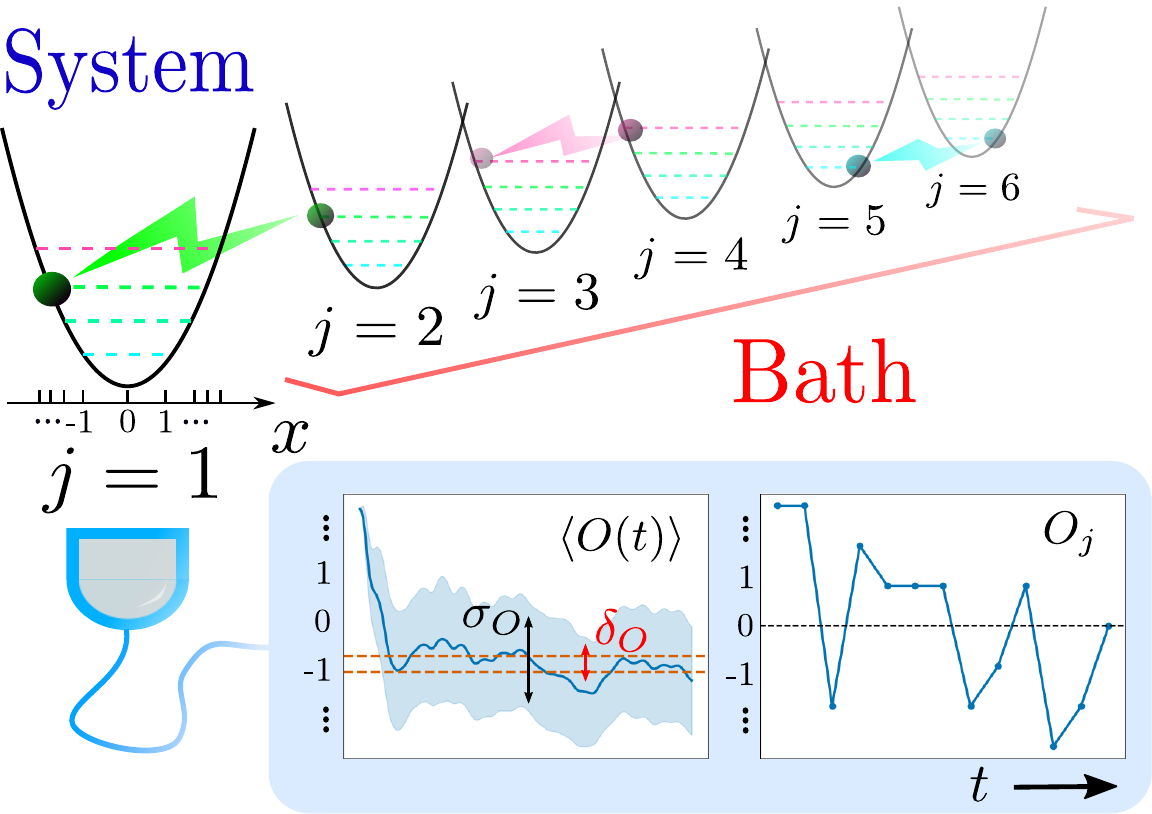}
	\caption{Diagram of general scheme. We consider systems with many energy levels, and thus many possible observable outcomes (diagram of Hamiltonian \eqref{eq:osc_Hamiltonian}). Local system observables may be measured via expectation value $\langle O(t) \rangle$ or via sequence of projective measurements $O_j$.}
	\label{fig:scheme} 
\end{figure}

We define the basis of eigenstates of $H_S$ and $H_B$,
\begin{equation}
\begin{split}
H_S | \epsilon_s \rangle &= \epsilon_s | \epsilon_s \rangle , \  s = 1, \dots, d_S , \\
H_B | E_{\beta}^{(B)} \rangle &= E_{\beta}^{(B)} | E_{\beta}^{(B)} \rangle, 
\  \beta = 1, \dots, d_B.
\end{split}
\end{equation} 
The (free) eigenstates of $H_0$ are $|\phi_\alpha\rangle$, with energy $E_\alpha$, and we define the index
$\alpha = 1, \dots, d_S d_B$, in order of increasing energy 
($E_{\alpha + 1} > E_\alpha$). 
Free eigenstates can be written as  
$|\phi_\alpha \rangle = |\epsilon_s \rangle | E_{\beta = f(\alpha, s)}^{(B)} \rangle$, 
with $f(\alpha,s)$ defined by the energy matching,
$
E_{f(\alpha,s)}^{(B)} = E_\alpha - \epsilon_s.
$
The (interacting) eigenstates of the total Hamiltonian $H$ are written as 
$|\psi_\mu\rangle = \sum_\alpha c_\mu(\alpha) |\phi_\alpha \rangle$. 
The total and bath DOS at energy $E$ are $D(E)$ and $D_B(E)$, respectively. In the limit of large system sizes they are related via:
\begin{equation}
D(E_\alpha) = \sum_s D_B(E_\alpha - \epsilon_s), 
\end{equation}	
which essentially counts the number of states of the bath that match the energies of the system. 

Consider a local observable $O = O_S \otimes \mathbb{1}_B$. 
We define operator matrix elements in the interacting basis by subscripts 
$\mu, \nu$, $O_{\mu\nu} := \langle \psi_\mu|O|\psi_\nu\rangle$, and free basis by subscripts
$\alpha, \beta$,  $O_{\alpha\beta} := \langle \phi_\alpha|O|\phi_\beta\rangle$. 
Non-interacting matrix elements can, in turn, be written in terms of local and bath degrees of freedom like 
$\langle \phi_\alpha |O|\phi_\beta\rangle = (O_S)_{s(\alpha)s(\beta)} \delta_{\alpha_B(\alpha)\alpha_B(\beta)}$,
where $s(\alpha)$ and $\alpha_B(\alpha)$ are the system and bath quantum numbers, respectively, of the free eigenstate $\alpha$.

\section{Chaotic Wavefunctions and Random Matrix Theory}
Studying the local dynamics of a generic many-body systems is a challenging task, and thus in order to treat such systems we make an ansatz on the generic structure of the eigenstates of such systems: that is, they take the form of \emph{quantum chaotic wavefunctions}. This may be defined by probability distribution over the coefficients $c_\mu(\alpha)$ of a many-body eigenstate $|\psi_\mu\rangle= \sum_\alpha c_\mu(\alpha)|\phi_\alpha\rangle$,
\begin{equation}\label{eq:prob_dist_c}
p(c) = \frac{1}{Z}\exp\left( -\sum_{\mu\alpha}\frac{c^2_\mu(\alpha)}{2\Lambda(\mu, \alpha)} \right)\prod_{\substack{\mu\nu\\ \mu>\nu}} \delta\left(\sum_\alpha c_\mu(\alpha)c_\nu(\alpha) \right).
\end{equation}
This probability distribution is simply Gaussian, except for the factor $ \delta\left(\sum_\alpha c_\mu(\alpha)c_\nu(\alpha) \right)$, which restricts the wavefunctions $|\psi_\mu\rangle$ such that they are mutually orthogonal \cite{Nation2018}. This description can be seen to be equivalent to a coarse-graining of the eigenstates, where the function $\Lambda(\mu, \alpha)$ describes the `envelope' of the chaotic wavefunctions. This ansatz has been directly numerically studied and shown to hold upon introduction of a non-integrable perturbation in realistic spin-chain systems in Refs. \cite{Atlas2017, Thesis}. Indeed, the transition to this Gaussian behaviour is seen to occur concurrently with the transition to Wigner-Dyson statistics of energy levels - a more typical marker of chaos in quantum systems \cite{DAlessio2016}.

The chaotic wavefunction approach thus depends on knowledge of the function $\Lambda$ describing the shape of the eigenstates of the model, which in general is a not obtainable analytically. Our approach here involves a drastic simplification, namely, we assume that $V$ is a real symmetric random matrix. The coarse-graining may then be taken as an average over realizations of the random perturbation (which may be understood as equivalent to an average over nearby energy levels \cite{Srednicki1994}). This assumption directly leads to the ETH, and also to effects that have been thoroughly checked in numerics in a variety of non-integrable systems \cite{Santos2010, Torres-Herrera2016, Borgonovi2016, Reimann2016, Nation2018}.
Formally, we express the matrix elements of $V$ in the free basis as random Gaussian numbers with  
$\langle V_{\alpha\beta} \rangle_V = 0$ and 
$\langle V_{\alpha\beta}^2 \rangle_V = \frac{g^2(1 + \delta_{\alpha\beta})}{N}$, 
where $\langle \cdots \rangle_V$ indicates an ensemble average over realizations of $V$. 
Furthermore, we assume that
$(H_0)_{\alpha\beta} = \alpha\omega_0 \delta_{\alpha\beta}$, 
with $\omega_0 = 1/N$. 
This last approximation only involves neglecting the variations in DOS within a relevant energy width (to be properly defined below).

The eigenstates of $H$ can be shown to follow a Lorentzian distribution \cite{Deutsch1991, Deutscha},
\begin{equation}\label{eq:Lambda}
\langle c_\mu^2(\alpha)\rangle_V := \Lambda(\mu, \alpha) = \frac{\omega_0 \Gamma / \pi}{(E_\mu - E_\alpha)^2 + \Gamma^2},
\end{equation}
with $\Gamma = \frac{\pi g^2}{N\omega_0}$ \cite{Nation2018}. We note that the Lorentzian form above is the only aspect relying on the random matrix model for an analytical foundation. Further, we note that for the two-body random interaction model, a more physically well justified random matrix approach, it is known that the function $\Lambda$ also takes a Lorentzian form (in which case it is usually referred to as the strength function) \cite{Flambaum1998, Borgonovi2016, Torres-Herrera2015}. We outline additional details of the chaotic wavefunction approach in Appendix \ref{App:SumRMT}, and a more thorough discussion is given in Ref. \cite{Thesis}.

In Ref. \cite{Nation2018} the current authors showed that this model leads to observable matrix elements, 
$O_{\mu\nu}$, in agreement with the ETH ansatz \cite{Srednicki1994, Srednicki1996}. 
This is achieved using a statistical theory of eigenstate correlation functions 
$\langle c_\mu(\alpha)c_\nu(\beta)\cdots\rangle_V$. 
Our model of chaotic wavefunctions can be shown to be self-averaging \cite{Nation2019a}, and thus taking the ensemble average to obtain such correlation functions is justified. See Appendix \ref{App:SumRMT} for technical details.

Continuing, we assume that the initial state for the quantum quench is an eigenstate of $H_0$, 
$| \psi(0) \rangle = |\phi_{\alpha_0}\rangle$, with eigenenergy $E_{\alpha_0}$,
though the formalism is easily extended to more general cases \cite{Nation2019}.  
We focus local observables that are diagonal in the free basis, 
$O_{\alpha\beta} \propto \delta_{\alpha\beta}$.
Our RMT model assumes a constant DOS, $1/\omega_0$, and coupling, $g$, leading also to a quantum chaotic eigenfunction width, $\Gamma$, that is independent of the energy. This theory can be applied to a generic quantum many-body system by the substitution 
$1/\omega_0 \to D(E_{\alpha_0})$. The RMT predicitions are valid as long as variations of $D(E)$ over the typical energy width $\Gamma$ can be neglected \cite{Nation2019a}. 

The main result of our previous work \cite{Nation2019} was an equation for the thermalization dynamics of an observable $O$, 
\begin{equation}\label{eq:time_evol}
\langle O(t) \rangle = (\langle O(t) \rangle_0 - \langle O(\infty) \rangle)e^{-2\Gamma t} + 
\langle O(\infty) \rangle,
\end{equation}
with the additional equality $\langle O(\infty) \rangle = \overline{[O_{\alpha\alpha}]}_{\alpha_0}$, and 
\begin{equation}\label{eq:micro_ave}
\overline{[O_{\alpha \alpha}]}_{\alpha_0} := \sum_\alpha \Lambda(\alpha_0, \alpha) O_{\alpha\alpha},
\end{equation}
is a microcanonical average of $O$ around the initial state energy $\alpha_0$. 
$\overline{[O_{\alpha \alpha}]}_{\alpha_0}$ can be physically understood as an average over the set of free eigenstates that are involved in the time evolution of the system. $\langle O(t) \rangle_0$ represents the free dynamics under $H_0$.

We now wish to study the time-averaged variance, or quantum fluctuations, of the local observable $O$,
$\sigma_O^2(\infty) = \langle O^2 (\infty) \rangle - \langle O (\infty) \rangle^2$, which can be obtained from Eq. \eqref{eq:time_evol} applied to $O$ and $O^2$,
\begin{equation}\label{eq:var1}
\sigma_O^2(\infty) =  \overline{[\Delta O^2_{\alpha \alpha}]}_{\alpha_0},
\end{equation}
where
$\overline{[\Delta O^2_{\alpha \alpha}]}_{\alpha_0} :=
\overline{[O^2_{\alpha \alpha}]}_{\alpha_0} - 
\overline{[O_{\alpha \alpha}]}_{\alpha_0}^2$.
%This result acquires an additional meaning in view of a
We recall a further result obtained in Ref. \cite{Nation2019}: the time-fluctuations of $O$ may be written as
\begin{equation}\label{eq:QCFDT}
\delta_O^2(\infty) = \frac{\overline{[\Delta O^2]}_{\alpha_0}}{4\pi D(E_{\alpha_0}) \Gamma}.
\end{equation}

From Eq. \eqref{eq:var1} and \eqref{eq:QCFDT} we may already observe a remarkable feature of fluctuations of chaotic systems, that is, their ratio after equilibration is given by,
\begin{equation}\label{eq:DOS_flucs}
\frac{\sigma_O^2(\infty)}{\delta_O^2(\infty)} = 4 \pi D(E_{\alpha_0}) \Gamma.
\end{equation}
The relative sizes of each of the fluctuation types has been previously understood from the ETH approach \cite{Srednicki1999}, however here we obtain both the precise scaling, and numerical prefactor. Eq. \eqref{eq:DOS_flucs} is our first relevant result, and may be understood as a signature of quantum ergodicity in many-body systems, and further reveals the DOS in terms of only measurable quantities (see Appendix \ref{App:MeasureDOS}). 

\section{Einstein Relation}
Now we show that Eq. \eqref{eq:var1} leads to the Einstein relation for the diffusion constant \cite{Kubo1966} in the limit $d_B \gg d_S \gg 1$, that is, a large system Hilbert space dimension.
To observe this, we re-express $\overline{[O_{\alpha \alpha}]}_{\alpha_0}$ via,
\begin{equation}\label{eq:prob_dist}
\overline{[O_{\alpha \alpha}]}_{\alpha_0}  = \sum_{s=-d_S/2}^{d_S/2} (O_S)_{s s} p(s),
\end{equation}
where $p(s)$ may be written in terms of the DOS of the bath (see Appendix \ref{App:ps}),
\begin{equation}\label{eq:p_s}
p(s) = \frac{D_B(E_{\alpha_0}-\epsilon_s)}{\sum_{s=-d_S/2}^{d_S/2} D_B(E_{\alpha_0}-\epsilon_s)}.
\end{equation}
To obtain the Einstein relation, we write
$D_B(E) = D_0 \exp( \beta(E) E))$
where $\beta(E)$ is the inverse microcanonical temperature, which we assume changes slowly over the width $\Gamma$. To make a connection with classical Brownian motion we consider now 
$O_S = X$, with $X_{s s'} =  s  \delta_{s,s'}$ and $ \epsilon_s = \frac{1}{2}m s^2$, interpreting the local quantum number $s$ as the position in an harmonic oscillator potential. 
In the limit of small temperature relative to the system bandwidth, and large compared to the system energy spacing, 
$1 \ll m\beta \ll d_S$,
we obtain,
\begin{equation}\label{eq:fdt_OU}
\sigma^2_X (\infty) = \overline{[X^2_{\alpha \alpha}]}_{\alpha_0} = \frac{1}{m \beta(E_{\alpha_0})}.
\end{equation}
Since $k_B T(E_{\alpha_0}) = \beta(E_{\alpha_0})^{-1}$ is the microcanonical temperature, we recover here the linear relation between the variance of the particle coordinate and the temperature that is found in OU processes (see Appendix \ref{App:OU}). Further, we note that Eq. \eqref{eq:fdt_OU} is an equipartition theorem, relating the average energy $\frac{1}{2}m\sigma^2_X (\infty)$ to the temperature. This occurs at the level of individual eigenstate averages $\overline{[X^2_{\alpha \alpha}]}_{\alpha_0}$, which motivates the description as an `eigenstate equipartition theorem'. In this sense, temperature can be defined not as a property of ensembles of systems, but rather as a property of individual chaotic eigenstates \cite{Borgonovi2017}.

We note that whilst they do not appear explicitly, typicality approaches are also able to produce similar fluctuation-relations \cite{Goldstein2006, Popescu2006}, however we stress important differences in the application and physical interpretation of the results. Firstly, our approach here is a \emph{dynamical} model, which allows for a non-equilibrium initial state, and hence Eq. \eqref{eq:fdt_OU} in the form $\sigma_X^2(\infty) = (m\beta(E_{\alpha_0}))^{-1}$ is a bonefide fluctuation-dissipation theorem relating the non-equilibrium decay rate to the equilibrium fluctuations that may not be understood from the typicality approach. 
Typicality instead describes the behaviour of `typical' states selected form a uniform distribution, and similar relations to the eigenstate equipartition theorem $\overline{[X^2_{\alpha \alpha}]}_{\alpha_0} = \frac{1}{m \beta(E_{\alpha_0})}$, can be inferred from such an approach. However, we here observe and justify their emergence for eigenstates of a single realization of a closed system, rather than `typical' states selected from a uniform distribution.

\section{Quantum Jump Trajectories}
We turn now to the case in which we perform a set of subsequent quantum measurements, and  assume that a non-degenerate local operator is measured. 
For the sake of clarity we consider again the operator $X$ defined above, an initial state
$|\psi(0)\rangle = |\epsilon_{s_0} \rangle |E^{(B)}_{\beta_0} \rangle $,
and a set of $N_m$ measurements separated by a time interval $\Delta t$, yielding a measurement record $s_1$, $s_2$, $\dots$, $s_{N_m}$.
The sequence of `measurement quenches' \cite{Bayat2018} is
\begin{equation}
|\epsilon_{s_0} \rangle | E^{(B)}_{\beta_0g} \rangle \to
|\epsilon_{s_1} \rangle | \psi^{(B)}_1 \rangle \to 
|\epsilon_{s_2} \rangle | \psi^{(B)}_2 \rangle \to 
\dots,
\end{equation}
where $|\psi^{(B)}_j\rangle$ is the state of the bath at step $j$. Assuming that the total energy is not significantly perturbed by the measurement process, the quantum dynamics is restricted to many-body states with energies close to the initial energy, 
$E_{\alpha_0} = \epsilon_{s_0} + E^{(B)}_\beta$. This assumption is valid assuming the range of system energies is negligible in comparison to the bath (see Appendix \ref{App:numerics} for a numerical validation of this assumption).

Eq. \eqref{eq:time_evol} is valid for any local observable and a different initial condition \cite{Nation2019}. 
We define $p(s_f,s_i;t_f,t_i)$ as the probability of measuring the value $s_f$ at time $t_f$, assuming that a previous observation yielded a value $s_i$ at time $t_i$. Thus, we can apply Eq. \eqref{eq:time_evol} to the projector $P_{s_f} = |\epsilon_{s_f} \rangle_S \langle \epsilon_{s_f}| \otimes \mathbb{1}_B$, and obtain
\begin{equation}
p(s_f,s_i;t_f,t_i) = \left( \delta_{s_f,s_i} - p_\infty(s_f) \right) e^{- 2 \Gamma \Delta t} + p_\infty(s_f),
\label{eq:markov_chain}
\end{equation}
where 
$p_\infty(s_f) = \overline{[(P_{s_f})_{\alpha \alpha}]}_{\alpha_0}$, and $\Delta t = t_f - t_i$. $p_\infty(s_f)$ is the steady-state probability for the system to be in state $s_f$, which in the RMT approach can be written in terms of a microcanonical ensemble around the initial energy $E_{\alpha_0}$. 

\begin{figure*}
		\includegraphics[width=\textwidth]{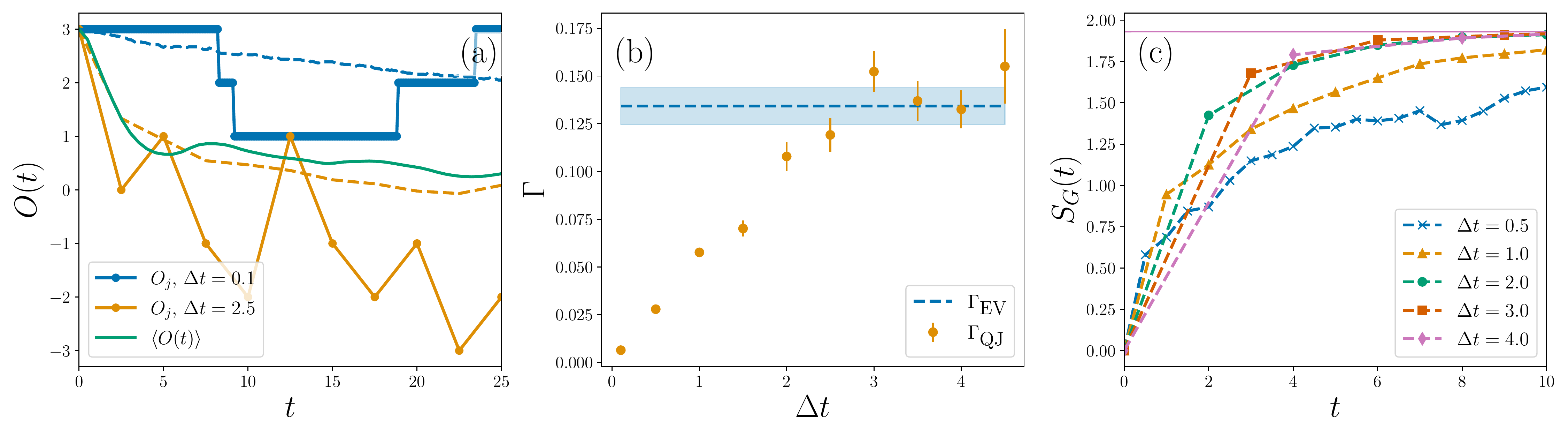}
	\caption{Exact diagonalization calculations of Hamiltonian \eqref{eq:osc_Hamiltonian}, \eqref{eq:QHO_coupling}. a) Observable dynamics as obtained from $\langle O(t)\rangle$ (green solid line), quantum jump trajectories $O_j$ (dotted lines), and their averages over 500 realizations (dashed lines). b) Convergence of the decay rate as measured by quantum jump trajectories ($\Gamma_{\textrm{QJ}}$) to that of thermalization dynamics ($\Gamma_{\textrm{EV}}$). Trajectories shown in Fig. \ref{fig:trajectories}. c) Growth of the non-equilibrium Gibbs entropy (Eq. \eqref{eq:Gibbs_ent}). Solid line shows single trajectory entropy for $\Delta t = 4$ (see below Eq. \eqref{eq:Gibbs_ent} for discussion). Parameters: $J = 0.8, h_x = 0.7, S = 3, N = 4$.}
	\label{fig:4plots} 
\end{figure*}

Eq. \eqref{eq:markov_chain} predicts that in the limit $\Delta t \gg 1/\Gamma$, the set of values $s_1$, $\dots$, $s_{N_m}$ will be scattered with variance $\sigma^2_X$. However, in the case $\Delta t < 1/\Gamma$, the measurement process may temporally resolve the decay of the initial value of $X$. Eq. \eqref{eq:markov_chain} in-fact predicts that the measurement outcomes form a Markov chain. Furthermore, we can show that the average over all the resulting stochastic trajectories of a measurement outcome, $s_j$ at time $t_j$, is the same as the expectation value $\langle X(t_j) \rangle$ at time $t_j$ (shown in Appendix \ref{App:QJT}). In other words, if we measure the expectation value $\langle X(t_j) \rangle$, the value is independent of whether we have subjected the systems to a quantum measurement at times $t < t_j$ or not. This is precisely the condition of `consistent histories' in \cite{Gemmer2014, Schmidtke2016}, which is conjectured as a mechanism behind the effective description of thermalization dynamics by a Markov process. Here we derive the mechanism from a description in terms of RMT, and more generally, chaotic wavefunctions. Finally, deviations from Eq. \eqref{eq:markov_chain} are expected for very short $\Delta t \ll t_Z$, with $t_Z$ a typical quantum Zeno time-scale. 

We may connect our discussion to the emergence of thermodynamic quantities through the non-equilibrium Gibbs entropy,
\begin{equation}\label{eq:Gibbs_ent}
S_G(t) = - \sum_{s-d_S/2}^{d_S/2} p(s,s_i;t,t_0)\ln p(s,s_i;t,t_0),
\end{equation}
and its behaviour with $\Delta t$. In fact, from Eq. \eqref{eq:time_evol} we may show that $\frac{d S_G(t)}{dt} \geq 0$ (see Appendix \ref{App:2ndLaw}). The definition of a non-equilibrium Gibbs entropy for quantum jump trajectories makes an important connection to results in stochastic thermodynamics \cite{Seifert2005, Seifert2012, Parrondo2015}. In particular, we have seen that $p(s,s_i;t,t_0)$ may be described by effective Langevin dynamics, and thus Eq. \eqref{eq:Gibbs_ent} may be seen to parallel the classical non-equilibrium Gibbs entropy defined in e.g. \cite{Seifert2005}. This construction further resembles the `observational entropy' in Refs. \cite{Safranek2019, Safranek2020}.

We finally note that an entropy may be defined for an \emph{individual} trajectory, by taking the probability distribution of measurement outcomes over all times. In equilibrium we have an equivalence between the quantum fluctuations $\sigma_O^2(\infty)$, and time-fluctuations of a single trajectory with $\Delta t \gg \Gamma^{-1}$, as each projective measurement occurs with a variance $\sigma_O^2(\infty)$. Thus, this entropy is equal to the maximal value of $S_G(t)$. This is confirmed numerically in Fig \ref{fig:4plots}c).

\section{Numerical calculations}
We have performed numerical experiments to check the validity of the RMT model and its predictions with two basic sets of models:

\subsection{Coupled quantum harmonic oscillators.} We consider a set of particles confined to move in a grid of discretized positions in one-dimensional harmonic potentials.
The Hilbert space is formed by states $|s,i\rangle$, where $s = -S, \cdots, S$ is the position in the $i^{\textrm{th}}$ potential,
\begin{equation}\label{eq:osc_Hamiltonian}
H_ 0 = \sum_{i=1}^N \sum_{s=-S}^S \epsilon_s |s,i \rangle \langle s,i| 
\end{equation}
with $\epsilon_s = s^2 $. To this, we add the coupling term
\begin{eqnarray}\label{eq:QHO_coupling}
& & V = h_x \sum_{i=1}^N \sum_{s=-S}^{S-1}(|s,i\rangle \langle s + 1,i| + H.c.)
+  \\
& & J \sum_{i=1}^{N-1}\sum_{s=-S}^{S-1}( |s,i\rangle \langle s+1,i+1| 
+ |s+1, i\rangle \langle s,i+1|  + H.c.)\nonumber ,
\end{eqnarray}
which includes both a kinetic energy term proportional to $h_x$, and a hopping $J$ between adjacent sites and energy levels in each oscillator. The observable is taken to be the oscillator position at $i=1$, $O = X_1 = \sum_s s|s, 1\rangle \langle s, 1|$.

Numerical results are shown in Fig. \ref{fig:4plots}. In particular, in Fig. \ref{fig:4plots}b) we see that the decay rate of averaged quantum jump trajectories indeed converges to that of $\langle O(t) \rangle$ outside of the Zeno regime. Further, we observe in Fig. \ref{fig:4plots}c) the growth of entropy in time to the value of the single trajectory entropy.

\subsection{Quantum Spin-Chains} 

The second system we consider is a Bilinear-Biquadtratic spin-chain \cite{Chubukov1991, Garcia-Ripoll2004, Lauchli2006}. Details and results are shown in the Appendix \ref{App:numerics}. In this case the Hamiltonian does not have a quadratic energy dispersion, an assumption only required for the comparison to the OU process. Further, we consider both a local and global observable of this model, finding that our analysis is valid in each case - our assumptions simply require the observable has a sufficiently sparse structure in the free basis \cite{Nation2019}. Finally, the dynamics of this model shows multiple timescales, which are resolved by the dynamics of the quantum jump trajectories when $\Delta t$ is of the relevant scale. This may allow quantum jump trajectories to resolve such phenomena as prethermalization \cite{Mori2018}. 

\section{Conclusions}
In this work we have shown how a closed quantum system initialized in a pure state may reproduce a \emph{classical} temperature dependent fluctuation-dissipation theorem of Brownian motion. Specifically, we have reproduced the Einstein relation for the Ornstein-Uhlenbeck process. This result is a direct analytical observation of the emergence of classical statistical physics from unitary quantum dynamics.  Indeed, we similarly observe an `eigenstate equipartition theorem', and thus see that microcanonical temperature relations can be seen on the level of individual eigenstates, thus extending the intuition afforded by the ETH.
Our results apply directly to quantum jump trajectories induced by repeated quantum measurements, finding that the trajectory is similarly described by a classical OU process.

Further, we observe that the fluctuations of chaotic quantum systems may be exploited to accurately measure its density of states. 

Our calculations are based on a random matrix theoretic approach, and build on earlier works where the current authors have obtained an analytic description of the full time-dependent decay to equilibrium \cite{Nation2019}. The current work formalises an important consequence of this approach, the emergence of a description of the fluctuations of local observables in terms of a microcanonical temperature. This hints to a quantum foundation of classical statistical physics, as we see the important properties of this theory directly from the quantum dynamics of pure states. We have confirmed our results by a numerical exact diagonalization calculations on two model systems.

We acknowledge discussions with Edgar Rold\'an, and funding from project PGC2018-094792-B-I00  (MCIU/AEI/FEDER, UE), EPSRC grant no.
EP/M508172/1, and from COST Action CA17113.

\appendix

\section{Summary of RMT Formalism}\label{App:SumRMT}

In this section we outline in brief the RMT methodology developed in Refs. \cite{Nation2018, Nation2019, Nation2019a, Thesis}, on which our calculations are based. We focus here on making clear the required assumptions on which the calculations rest, and refer the reader to the above references for details on the calculations themselves. Ref. \cite{Nation2018} provides a detailed formulation of the RMT model, and a derivation of the ETH, Ref. \cite{Nation2019} extends and formalises key features of observables, and describes time evolution of observables, and Ref. \cite{Nation2019a} extends the approach to finite temperatures, and applies the method to an application on quantum computers and other devices. Self-averaging of chaotic wavefunctions is shown and discussed in the appendices of Ref. \cite{Nation2019a}. The thesis \cite{Thesis} goes into more detail regarding the assumptions on observables made below, obtaining physical conditions for the fulfilment of the crucial assumptions. Each of these works includes exact numerical calculations of realistic quantum spin-chains, which compare very well with the RMT framework.

Our summary below will be separated into two sections, the assumptions required on chaotic wavefunctions, and those on observables.

\subsection{Assumptions on chaotic wavefunctions}

The main assumption of our RMT formalism is the ansatz that the probability distribution on chaotic wavefunctions, $|\psi_\mu\rangle = \sum_\alpha c_\mu(\alpha)|\phi_\alpha\rangle$, is a Gaussian distribution with the constraint of mutual orthogonality, $\langle \psi_\mu|\psi_\nu\rangle = \delta_{\mu\nu}$,
\begin{equation}\label{eq:c_prob_dist}
p(c) = \frac{1}{Z}\exp\left( -\sum_{\mu\alpha}\frac{c^2_\mu(\alpha)}{2\Lambda(\mu, \alpha)} \right)\prod_{\substack{\mu\nu\\ \mu>\nu}} \delta\left(\sum_\alpha c_\mu(\alpha)c_\nu(\alpha) \right).
\end{equation}
That is, the action of the interaction causes the eigenstate $|\psi_\mu \rangle$ to mix with sufficiently many approximately non-interacting states $|\phi_\alpha\rangle$ such that the distribution may be described by a Gaussian with some width $\Lambda(\mu, \alpha)$, with the requirement that the eigenstates remain orthogonal. The function $\Lambda$ thus yields the envelope of the random wavefunctions. This function is shown to be a Lorentzian of width $\Gamma$ for the particular RMT model which we use for comparison to our model, though it may be different for different models.
In general for chaotic systems one may expect this function to be peaked around a certain energy, with a width $\Gamma(E)$ that may depend on the energy of the wavefunction. We show in \cite{Nation2019a} that this change in width with energy can in fact be incorporated into our theory.

From Eq. \eqref{eq:c_prob_dist} one can calculate arbitrary correlation functions $\langle c_\mu(\alpha) \cdots c_\nu(\beta)\rangle_V$ of the model \cite{Nation2018}. We see that the largest correlation function that does not factorize is the four point correlation function,
\begin{equation}
\begin{split}
& \langle c_\mu(\alpha)c_\nu(\beta)c_\mu(\alpha^\prime)c_\nu(\beta^\prime)\rangle_V = \Lambda(\mu, \alpha)\Lambda(\nu, \beta)\delta_{\alpha\alpha^\prime}\delta_{\beta\beta^\prime} \\ & - \frac{\Lambda(\mu, \alpha)\Lambda(\nu, \beta)\Lambda(\mu, \alpha^\prime)\Lambda(\nu, \beta^\prime)}{ \sum_\alpha \Lambda(\mu, \alpha)\Lambda(\nu, \alpha)}(\delta_{\alpha\beta}\delta_{\alpha^\prime\beta^\prime} + \delta_{\alpha\beta^\prime}\delta_{\beta\alpha^\prime}).
\end{split}
\end{equation}
This can be understood in terms of Gaussian and non-Gaussian contractions, where the first term is that due to purely Gaussian behaviour (reminiscent of Wicks' theorem, for example), and the second term is due to the effective interactions between chaotic wavefunctions due to mutual orthogonality. We note that this term is actually crucial for a consistent description of observable matrix elements, and time-evolution.
It is these correlation functions that form the basis for calculations in our framework.

\subsection{Assumptions on observables}\label{sec:obs}

For the work outlined above there are two relevant assumptions to be made on the form of observables. The first, is that we assume that in the non-interacting basis the observable is diagonal, so $O_{\alpha\beta} \propto \delta_{\alpha\beta}$. We note that this is not a requirement for the general framework, which can be extended to observables that take instead a sparse structure in this basis \cite{Nation2019}.

The second assumption can be summarized as `the ability to define a microcanonical average that does not vary pathologically in energy'. We will detail the specific requirements for this below, but note that this can be understood simply to be a minimal requirement on observables in order for thermalization to occur, as thermalization requires that a system observables evolve to a microcanonical state that does not depend on the particular microstate of the initial state, rather on its energy alone.

\begin{figure*}
	\includegraphics[width=0.65\textwidth]{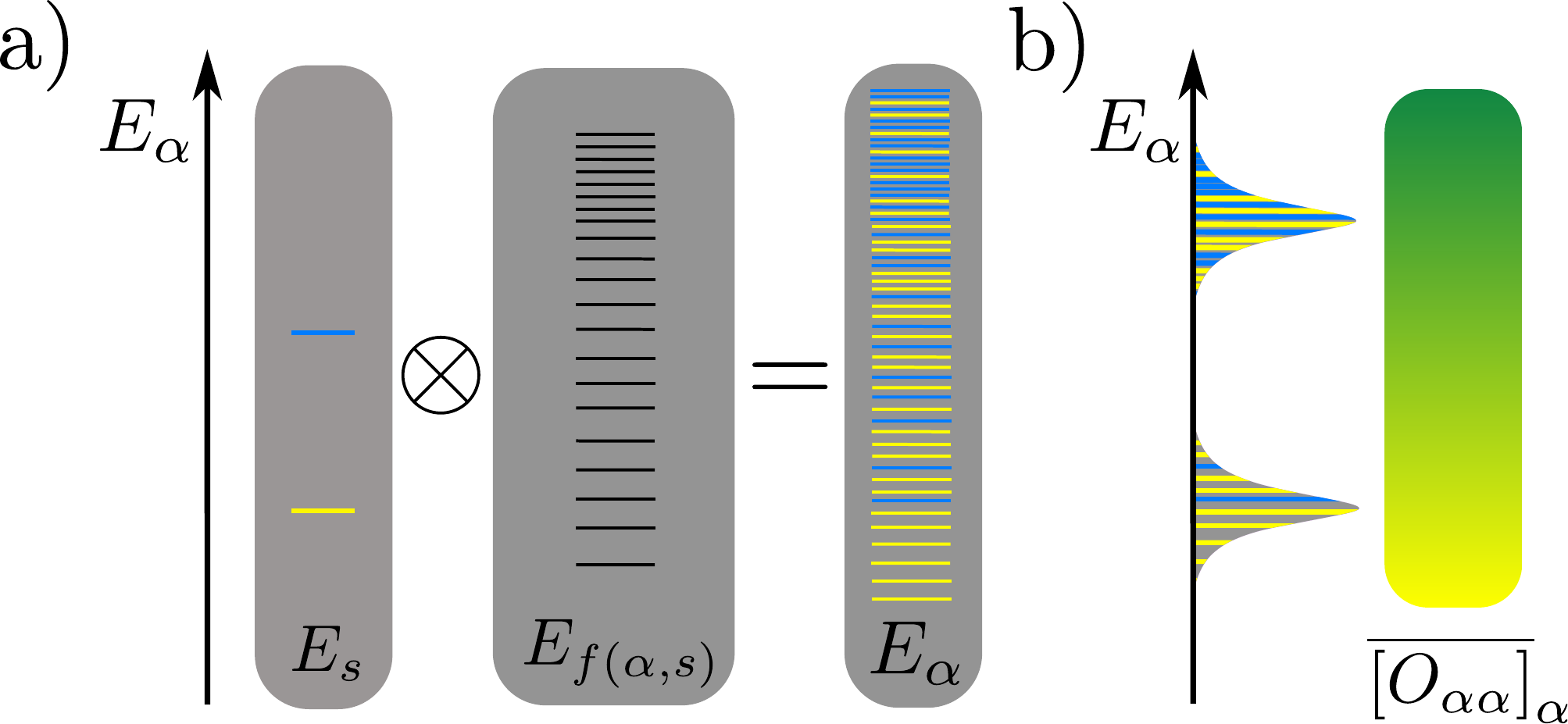}
	\caption{a) Energy level diagram. Shown are two system energy levels $E_s$ in distinct colors, the bath energy levels $E_{f(\alpha, s)}$, and the total system + bath energy levels $E_\alpha$, coloured according to their respective system state. b) Illustration of microcanonical average $\overline{[O_{\alpha\alpha}]}_\alpha$. Each level in the average is weighted by the function $\Lambda$, this average is assumed to be made up of many energy levels, and to vary smoothly with energy.}
	\label{fig:mc_averages} 
\end{figure*}

In detail, then, this assumption requires that the microcanonical average
\begin{equation}
\overline{[O_{\alpha\alpha}]}_\mu := \sum_\alpha \Lambda(\mu, \alpha)O_{\alpha\alpha},
\end{equation}
is smooth over the width $\Gamma$ of the function $\Lambda$. This is illustrated in Fig. \ref{fig:mc_averages}. We showed in Ref. \cite{Nation2019} that this smoothness condition is fulfilled under the two conditions:
\begin{equation}
\begin{split} 
&\frac{\Gamma}{\omega_0} \gg 1 \\
\Gamma^2 & \left|\frac{d^2 \overline{[O_{\alpha\alpha}]}_\mu}{dE_\mu^2}\right| \ll 1.
\end{split}
\end{equation}

\section{Derivation of Eq. (9)}\label{App:ps}

In this section we will evaluate 
\begin{equation}\label{eq:micro_aveS}
\overline{[O_{\alpha \alpha}]}_{\alpha_0} := \sum_\alpha \Lambda(\alpha_0, \alpha) O_{\alpha\alpha}.
\end{equation}
The important point here is to realize that the $\Lambda$ functions in the sum in Eq. \eqref{eq:micro_aveS}, which are Lorentzian distributions of width $\Gamma$, can be approximated as delta-functions (for small enough values of $\Gamma$); explicitly selecting those values such that $E_\alpha = E_{\alpha_0}$ in the summation,
\begin{equation}
\overline{[O_{\alpha \alpha}]}_{\alpha_0} = 
\sum_\alpha O_{\alpha \alpha} 
\frac{1}{D(E_{\alpha_0})}
\frac{\Gamma_{\alpha_0}/\pi}{(E_\alpha - E_{\alpha_0})^2 + \Gamma^2_{\alpha_0}} .
\end{equation}
Note that $E_{\alpha_0}$ and $E_\alpha$ can be interchanged in the definition of $\Lambda$, since we require that both $\Gamma_{\alpha}$ and $D(E_\alpha)$ vary negligibly over energy scales of the order of $\Gamma_{\alpha}$. Under this very approximation we can change the Lorentzian by a Dirac delta function. Additionally, we work in the continuum limit, such that we may re-express the sum over $\alpha_B$ as an integral over the bath eigenstates, $\sum_{\alpha_B} \to \int dE_{\alpha_B} D_B(E_{\alpha_B})$. We thus have,
\begin{equation}
\begin{split}
& \overline{[O_{\alpha \alpha}]}_{\alpha_0} =
\sum_s O_{s s} \sum_{\alpha_B}  
\frac{1}{D(E_{\alpha_0})} \delta \left( E_\alpha - E_{\alpha_0} \right) \\
&=
\sum_s O_{s s} \int d E_{\alpha_B} 
\frac{D_B(E_{\alpha_B})}{D(E_{\alpha_0})} \delta \left( E_\alpha - E_{\alpha_0} \right) \\
&=
\sum_s O_{s s} \int d E_{\alpha_B}   
\frac{D_B(E_{\alpha_B})}{D(E_{\alpha_0})} \delta \left( \epsilon_s + E_{\alpha_B} - E_{\alpha_0} \right)  \\
&=
\sum_s O_{s s} D_B(E_{\alpha_0}-\epsilon_s)  \frac{1}{D(E_{\alpha_0})}  \\
&=
\sum_s O_{s s} p(s).
\end{split}
\end{equation}
Here we have defined the probabilities
\begin{eqnarray}
p(s) = \frac{D_B(E_{\alpha_0}-\epsilon_s)}{D(E_{\alpha_0})} = 
\frac{D_B(E_{\alpha_0}-\epsilon_s)}{\sum_s D_B(E_{\alpha_0}-\epsilon_s)}.
\end{eqnarray}
Notably, for the special case where the bath density of states  does not change over the entire system energy spectrum, we thus recover $p(s) = \frac{1}{d_S} \forall s$. This is a common assumption in formulations of statistical physics: that of {\it equal a-priori probabilities}. We thus observe the physical requirement for this common assumption of statistical physics to be valid within our theory.

\section{Derivation of Eq. (11)}\label{App:Temp_DOS}

In this section we show that 
\begin{equation}\label{eq:DeltaO_pS}
\begin{split}
\overline{[\Delta O^2_{\alpha \alpha}]}_{\alpha_0} &=
\sum_s p(s) O_{ss}^2 - \left( \sum_s p(s) O_{ss} \right)^2 \\& 
\sim \beta^{-1}, 
\end{split}
\end{equation}
for a system with a harmonic energy dispersion $E_s = \frac{1}{2}m s^2$.

In this case, we have the partition function
\begin{equation}
{\cal Z}(\beta) = \sum_{s} e^{-\beta E_s^2},
\end{equation}
where $s$ takes $2S + 1$ possible values from $[-S, S]$ (or more generally $d_S$ values from $[-\frac{d_S}{2}, \frac{d_S}{2}]$), and $\beta = \beta(E_\alpha)$. This can itself be evaluated as a Gaussian integral, $\sum_{s} \to \int_{-\infty}^\infty ds$, such that
\begin{equation}
\begin{split}
{\cal Z}(\beta^\prime) &= \int_{-\infty}^\infty ds e^{-\frac{1}{2}\beta^\prime s^2} \\& 
= \sqrt{\frac{2\pi}{\beta^\prime}},
\end{split}
\end{equation}
where we have defined $\beta^\prime := m\beta$.
Now, the first term in Eq. \eqref{eq:DeltaO_pS} can be written as
\begin{equation}
\begin{split}
\overline{[O_{\alpha\alpha}^2]}_{\alpha_0} &= \sum_s p(s) o_{ss}^2 \\&
= \frac{1}{Z(\beta^\prime)}\sum_s s^2 e^{-\frac{1}{2}\beta^\prime s^2} \\&
= \frac{1}{Z(\beta^\prime)}\int_{-\infty}^\infty ds s^2 e^{-\frac{1}{2}\beta^\prime s^2} \\& 
= \frac{1}{\beta^\prime}.
\end{split}
\end{equation}

Now, the second term in Eq. \eqref{eq:DeltaO_pS}, can be seen along the same lines to be trivially zero, 

we thus have,
\begin{equation}\label{eq:O2}
\overline{[O_{\alpha\alpha}^2]}_{\alpha_0} = \frac{1}{m \beta}.
\end{equation}

\begin{figure*}
	\includegraphics[width=\textwidth]{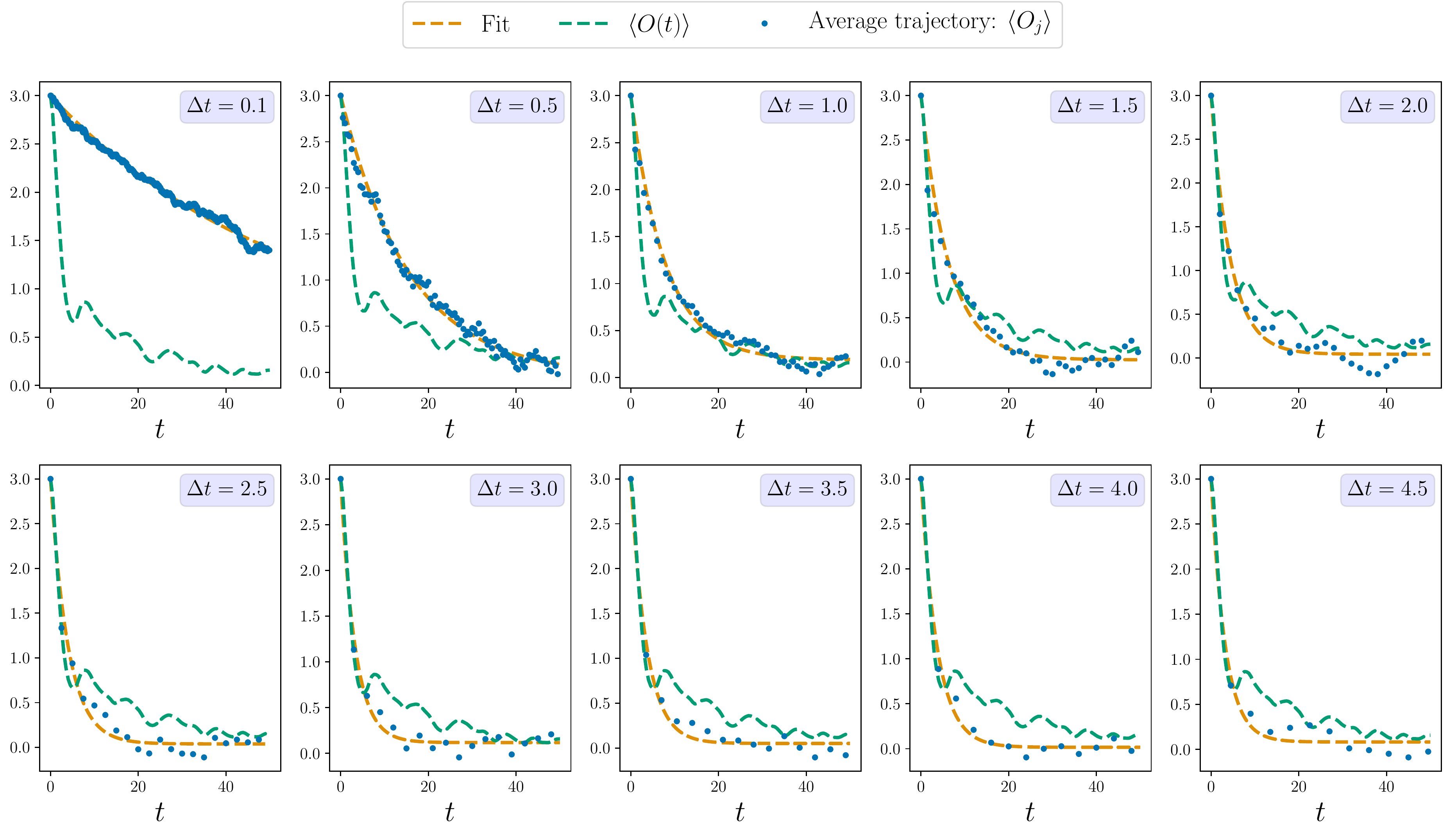}
	\caption{Average quantum jump trajectories of $O = X_1$ quantum harmonic oscillator Hamiltonian \eqref{eq:osc_Hamiltonian} and \eqref{eq:QHO_coupling} of the main text for increasing $\Delta t$ with a)-j). Here we see that as $\Delta t$ is increased the decay rate of the average jump trajectory decays at the same rate as the expectation value (green dashed line). Orange dashed line shows fit to exponential decay used to obtain $\Gamma_{\textrm{QJ}}$ in Fig. 2 b). For small $\Delta t$, the decay is slowed due to proximity to the Zeno regime of completely frozen dynamics at $\Delta t \to 0$. Averages over 500 realizations of quantum trajectories (100 realizations for $\Delta t = 0.1, 0.5$). Parameters: $J = 0.8, h_x = 0.7, S = 3, N = 4$. $\Delta t =$ 0.1, 0.5, 1, 1.5, 2., 2.5, 3, 3.5, 4, 4.5 for plots a)-j) respectively.}
	\label{fig:trajectories} 
\end{figure*}

\section{Quantum Jump Trajectories}\label{App:QJT}

\subsection{Thermalization of Quantum Jump Trajectories}

We can show that, according to expression (13), the probability distribution of measurement outcomes at a given time $t$ is independent of measurements having been performed at times between $0$ and $t$. 
This implies that the average over quantum jump trajectories of the measurement outcome of an observable, $O$, at some time $t$, is the same as the expectation value $\langle O (t) \rangle$ in the absence of previous quantum measurements. 

This can be shown with the following relation. Assume that a measurement yields a value $s_i$ at time $t_i$ and a future observation yields the value $s_f$ at time $t_f$. At some intermediate time, an observation is performed a time $t_i < t' < t_f$, with outcome $s'$. From simple algebra it follows that the conditioned probability distribution in (13) satisfies that,
\begin{equation}
\begin{split}
\sum_{s_m}&  p(s_f,s_m;t_f,t_m)p(s_m,s_i;t_m,t_i)\\& = 
\sum_{s_m}
\left( (\delta_{s_f,s_m} - p_{\infty}(s_f) ) e^{-2 \Gamma (t_f - t_m)} + p_\infty(s_f) \right) \\&
\qquad \times \left( (\delta_{s_m,s_i} - p_{\infty}(s_m) ) e^{-2 \Gamma (t_m - t_i)} + p_\infty(s_m) \right) 
\\&
= (\delta_{s_f,s_i} - p_{\infty}(s_f) ) e^{-2 \Gamma(t_f-t_i)} 
+ p_\infty(s_f) \\ &
= p(s_f,s_i;t_f,t_i).
\end{split}
\label{eq:markov}
\end{equation}

By induction Eq. \eqref{eq:markov} can be extended to the case where an average is taken over a set of intermediate measurement outcomes, yielding the result that the average distribution probability at some time is independent of whether the system was monitored or not. This result is of course not valid in the Zeno regime, where the exponential decay assumption is not valid.

\begin{figure*}
	\includegraphics[width=\textwidth]{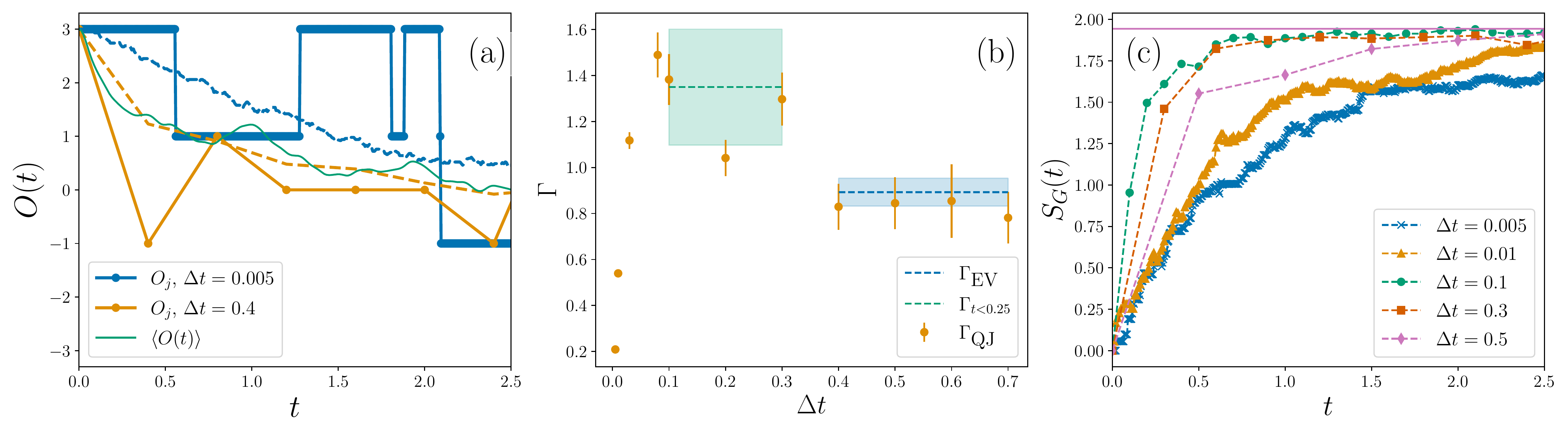}
	\caption{Exact diagonalization calculations of spin-chain Hamiltonian of Eq. \eqref{eq:SC_Hamiltonian} for $O = S_z^1$. a) Examples of observable dynamics as obtained from $\langle O(t)\rangle$, quantum jump trajectories $O_j$, and their averages over 100 realizations (dashed lines). b) Convergence of the decay rate as measured by quantum jump trajectories to that of thermalization dynamics. Trajectories shown in Fig. \ref{fig:SC_trajectories}. c) Growth of the non-equilibrium Gibbs entropy. Solid line shows single trajectory entropy for $\Delta t = 0.5$. Parameters: $N = 4, S = 3, h_z = 1, h_x  =0.2, J = 0.8, \Delta = 0.3, q = 1.5$.}
	\label{fig:3plots_SC} 
\end{figure*}
\begin{figure*}
	\includegraphics[width=\textwidth]{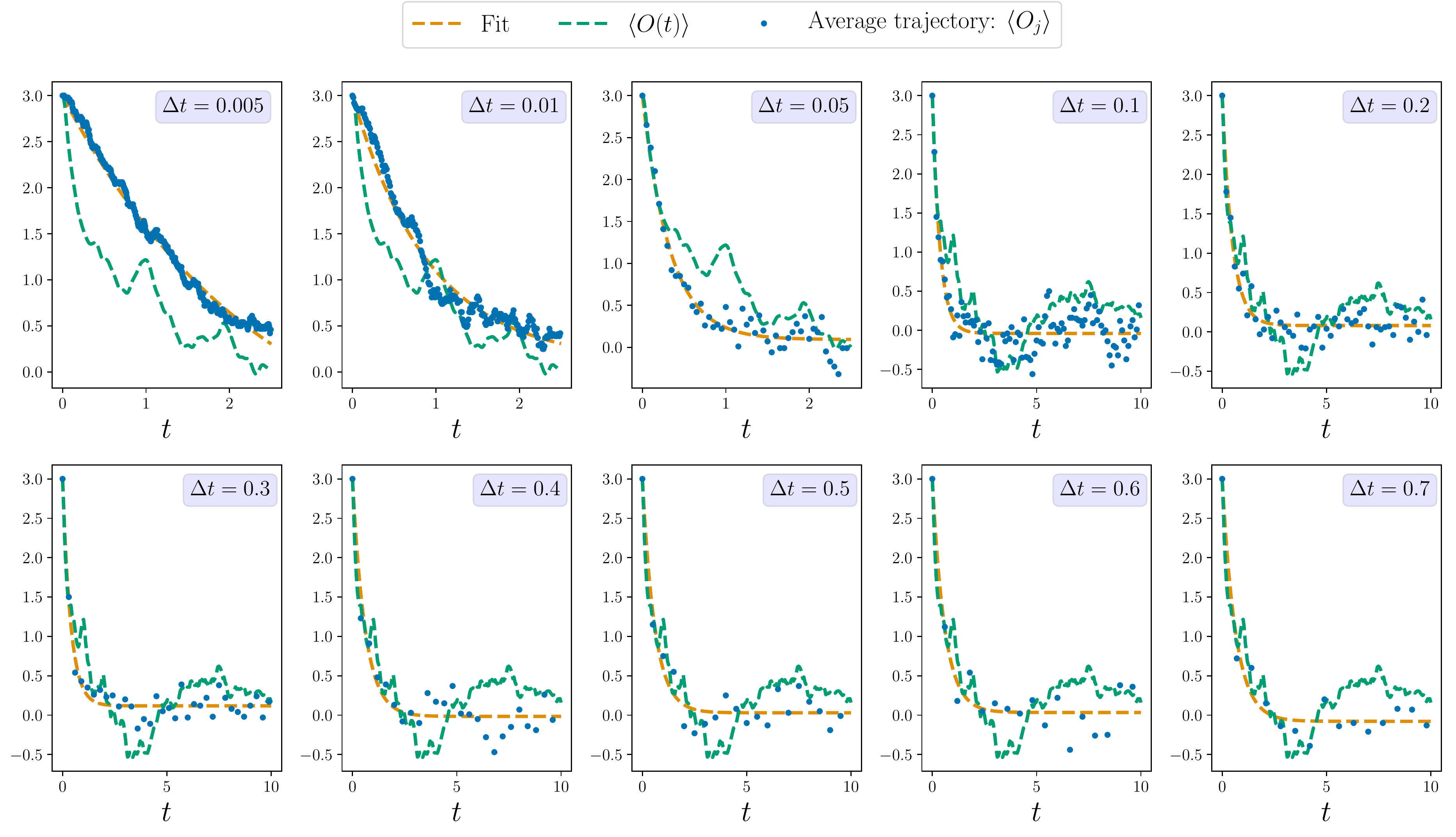}
	\caption{Average quantum jump trajectories of $O = S_z^1$ of spin-chain Hamiltonian of Eq. \eqref{eq:SC_Hamiltonian} for increasing $\Delta t$ with a)-j). Here we see that as $\Delta t$ is increased the decay rate of the average jump trajectory decays at the same rate as the expectation value (green dashed line). Orange dashed line shows fit to exponential decay used to obtain $\Gamma_{\textrm{QJ}}$ in Fig. \ref{fig:3plots_SC}b). For small $\Delta t$, the decay is slowed due to proximity to the Zeno regime of completely frozen dynamics at $\Delta t \to 0$. Averages over 100 realizations of quantum trajectories. Parameters: $N = 4, S = 3, h_z = 1, h_x  =0.2, J = 0.8, \Delta = 0.3, q = 1.5$. $\Delta t =$ 0.005, 0.01, 0.05, 0.1, 0.2, 0.3, 0.4, 0.5, 0.6, 0.7 for plots a)-j) respectively.}
	\label{fig:SC_trajectories} 
\end{figure*}

We show examples of the decay of average quantum jump trajectories in Fig. \ref{fig:trajectories}. These are the same trajectories used to obtain the decay rates in Fig. 2b) of the main text.

\subsection{Derivation of the $2^{\textrm{nd}}$ law}\label{App:2ndLaw}
In this section we will bound the derivative of the non-equilibrium Gibbs entropy. To simplify notation, we write $p(s_f,s_i;t_f,t_0) = p(s_f;t)$, such that
\begin{equation}
S_G(t) = - \sum_{s=-d_S/2}^{d_S/2} p(s, t)\ln p(s, t),
\end{equation}
for 
\begin{equation}
p(s, t) = p(s, 0)e^{-2\Gamma t} + (1 - e^{-2\Gamma t})p_\infty(s),
\end{equation}
where $p_\infty(s)$ is the equilibrium probability of obtaining the outcome $s$ from a measurement of $O$. We have, then, that
\begin{equation}
\begin{split}
\frac{d S_G(t)}{dt} &= - \sum_s [2\Gamma e^{-2\Gamma t} (p_\infty(s) - p(s, 0) ) (\ln p(s, t) + 1)] \\&
\end{split}
\end{equation}
which, using that $1 - \frac{1}{x} \leq \ln x \leq x - 1$, we obtain
\begin{equation}
\begin{split}
\frac{d S_G(t)}{dt} &\geq 2\Gamma e^{-2\Gamma t} [ \sum_s (p(s, 0) p(s, t) - p_\infty(s)(2 - \frac{1}{p(s, t)}] \\& 
\geq 2\Gamma e^{-2\Gamma t} [p(s_0, t) - 2 + \sum_s \frac{1}{p(s, t)}]
\end{split}
\end{equation}
where in the second line we have used that $p(s, 0) = \delta_{s, s_0}$, where $s_0$ is the initial value of $O_S$. Now, we can thus see that at $t \to 0$, the factor $\sum_s \frac{1}{p(s, t)} \to \infty$. This indicates that at early times the entropy grows faster for smaller $\Delta t$, as observed in Fig. 2c) of the main text. For $t > 0$, we can note that $\frac{1}{p(s, t)} \geq 1$ and $p(s_0, t) \geq 0$, so 
\begin{equation}
\begin{split}
\frac{d S_G(t)}{dt} &\geq 2\Gamma e^{-2\Gamma t} [d_s - 2] > 0,
\end{split}
\end{equation}
for observables with more than one possible outcome $d_s \geq 2$.

Interpreting this result, we note that $S_G(t)$ is defined for quantum jump trajectories only at times $j \Delta t$, and we have that $p(s, t)$ follows the RMT result between successive measurements. We thus see that, averaged over trajectories, the Gibbs entropy can be seen to increase between successive measurements.

\section{Additional Numerical Results}\label{App:numerics}

In this section we present some complementary numerical results to the results of the main text. Firstly, we present in Fig. \ref{fig:trajectories} the corresponding quantum jump trajectories to the decay rate plot of Fig. 2b) of the main text. These show the decay of the expectation value, as well as the quantum jump trajectories for different values of $\Delta t$, to which we perform a fit. Notice that for $\Delta t$ outside of the Zeno regime, we observe the quantum jump trajectories thermalize at approximately the same rate as the expectation value.

In each case, we initialize the system a mid-energy eigenstate of the non-interacting Hamiltonian, $H_0$, choosing such that $\langle O(0)\rangle = \max(O)$, and obtain $\Gamma$ from a fit to Eq. \eqref{eq:time_evol}.

\subsection{Quantum Spin Chain Results}

In Fig. \ref{fig:3plots_SC} we show complementary results to Fig. 2 of the main text for the large $S$ spin-chain given by the Hamiltonian
\begin{equation}\label{eq:SC_Hamiltonian}
H_0 = \sum_j^N \bigg[ h_z S_z^{j} + h_x S_x^{j} \bigg],
\end{equation}
where $j=1$ is the system spin. The coupling Hamiltonian is
\begin{equation}
\begin{split}
V &= \frac{1}{2}J\sum_i^{N-1}\bigg[ S_x^{i}S_x^{i+1} + S_y^{i}S_y^{i+1} + \Delta S_z^{i}S_z^{i+1} 
\\& + q\big( (S_x^{i}S_x^{i+1})^2 + (S_y^{i}S_y^{i+1})^2 + \Delta (S_z^{i}S_z^{i+1})^2 \big) + H.c \bigg],
\end{split}
\end{equation}
where $S_{x, y, z}^{i}$ are spin operators on site $i$. Notice that this Hamiltonian does not have a quadratic energy dispersion of the system at $i = 1$ - this is required only to obtain the Einstein relation in the form of the OU process.

The contributing thermalization dynamics of both the expectation values and quantum jump trajectories, used to obtain Fig. \ref{fig:3plots_SC}b), are shown in Fig. \ref{fig:SC_trajectories}. Here we have used the observable $O = S_z^{1}$.

Interestingly, we observe that the expectation value dynamics consist of two separate timescales. At very short times, the decay is fast, however after some time, a slower decay dominates. Notice that this more complicated dynamics is mirrored in the quantum jump trajectories. In Fig. \ref{fig:3plots_SC}b), unlike the harmonic oscillator chain, the quantum jump trajectory decay rate is actually faster than the expectation value decay for a range of $\Delta t$. For this intermediate range of $\Delta t$ values, the quantum jump trajectories decay at the same rate as the short time dynamics of the expectation value. As $\Delta t$ is increased, the decay rate slows to that of a fit to the whole dynamics of the trajectory. 

We thus see that more complex dynamics may also be resolved in the quantum jumps framework. Indeed, the approach from quantum chaos, employing Eq. \eqref{eq:c_prob_dist}, is more general than the specific RMT model applied in the main text, and may describe systems where $\Lambda$ is of a different form to a Lorentzian. In such cases, the decay deviates from a purely exponential form.

\subsubsection{Global Observables}

\begin{figure*}
	\includegraphics[width=\textwidth]{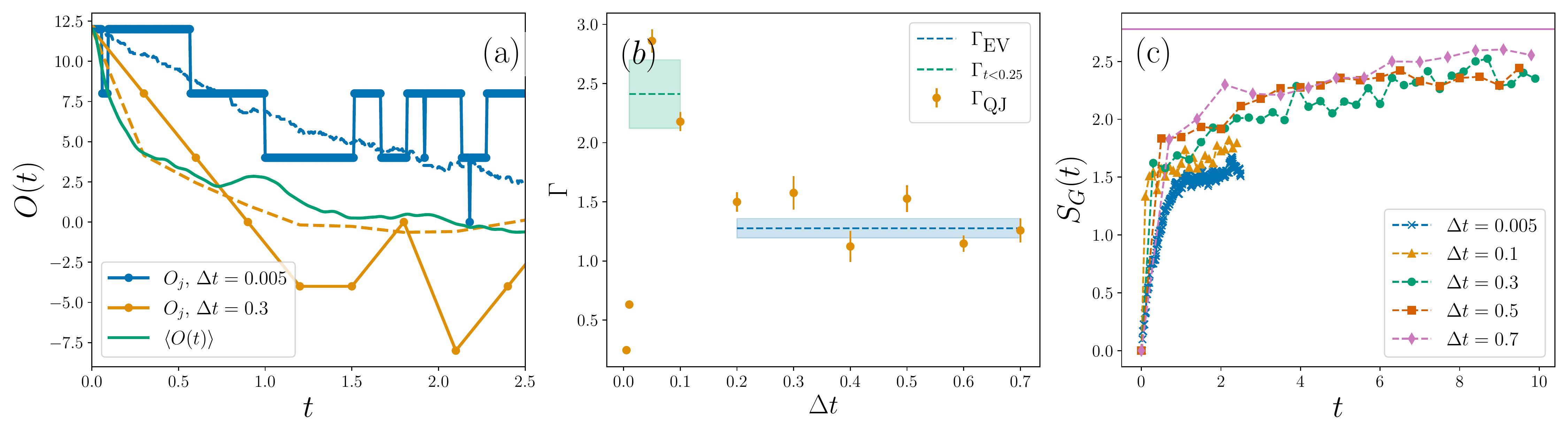}
	\caption{Exact diagonalization calculations spin-chain Hamiltonian of Eq. \eqref{eq:SC_Hamiltonian} for global observable $O = \sum_i^N S_z^i$. a) Examples of observable dynamics as obtained from $\langle O(t)\rangle$, quantum jump trajectories $O_j$, and their averages over 100 realizations, indicated by $\langle O_j \rangle$ (dotted lines). b) Convergence of the decay rate as measured by quantum jump trajectories to that of thermalization dynamics. Trajectories shown in Fig. \ref{fig:Global_trajectories}. c) Growth of the non-equilibrium Gibbs entropy. Solid line shows single trajectory entropy for $\Delta t = 0.7$. Parameters: $N = 4, S = 3, h_z = 1, h_x  =0.2, J = 0.8, \Delta = 0.3, q = 1.5$.}
	\label{fig:3plots_Global} 
\end{figure*}

\begin{figure*}
	\includegraphics[width=\textwidth]{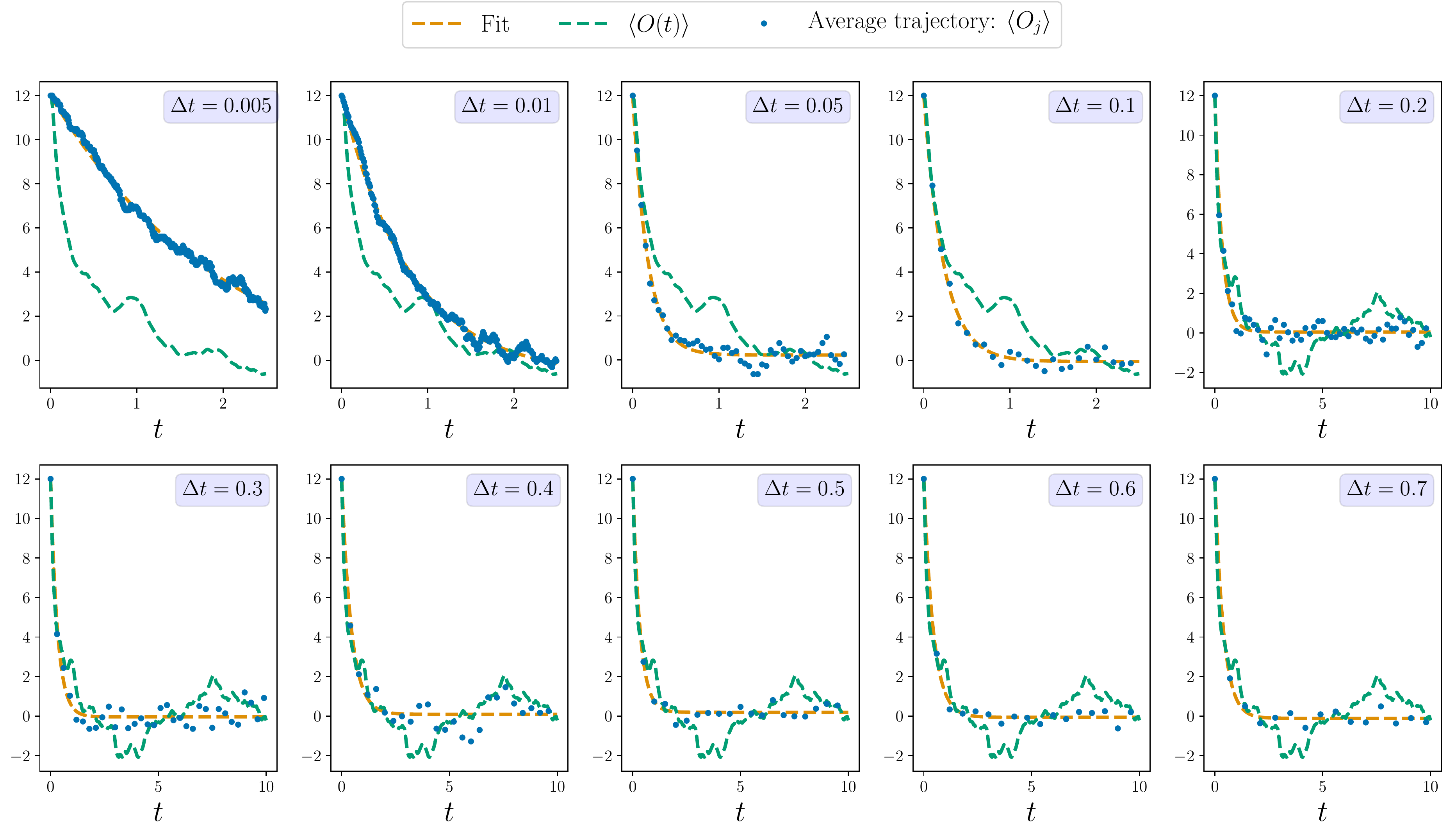}
	\caption{Average quantum jump trajectories of $O = \sum_i S_z^{i}$ of spin-chain Hamiltonian of Eq. \eqref{eq:SC_Hamiltonian} for increasing $\Delta t$ with a)-j). Here we see that as $\Delta t$ is increased the decay rate of the average jump trajectory decays at the same rate as the expectation value (green dashed line). Orange dashed line shows fit to exponential decay used to obtain $\Gamma_{\textrm{QJ}}$ in Fig. \ref{fig:3plots_Global}b). For small $\Delta t$, the decay is slowed due to proximity to the Zeno regime of completely frozen dynamics at $\Delta t \to 0$. Averages over 100 realizations of quantum trajectories. Parameters: $N = 4, S = 3, h_z = 1, h_x  =0.2, J = 0.8, \Delta = 0.3, q = 1.5$. $\Delta t =$ 0.005, 0.01, 0.05, 0.1, 0.2, 0.3, 0.4, 0.5, 0.6, 0.7 for plots a)-j) respectively.}
	\label{fig:Global_trajectories} 
\end{figure*}

The theory developed in the main text does not require that the observable is strictly local, rather that is diagonal in the basis of eigenstates of the non-interacting Hamiltonian. In-fact, even this requirement is not necessary in our RMT framework, rather the observable must be sufficiently sparse, and may be formulated in terms of sums of local observables $O = \sum_i O_i$ that are not necessarily diagonal \cite{Nation2019}.

We can thus apply this approach to global observables of the system. Here we use the spin-chain system of Eq. \eqref{eq:SC_Hamiltonian}, and choose as our observable $O = \sum_j S_z^{j}$. We see in Figs. \ref{fig:3plots_Global} and \ref{fig:Global_trajectories} that out analysis of the main text still holds in this case.

\subsection{Total Energy}

\begin{figure*}
	\includegraphics[width=0.32\textwidth]{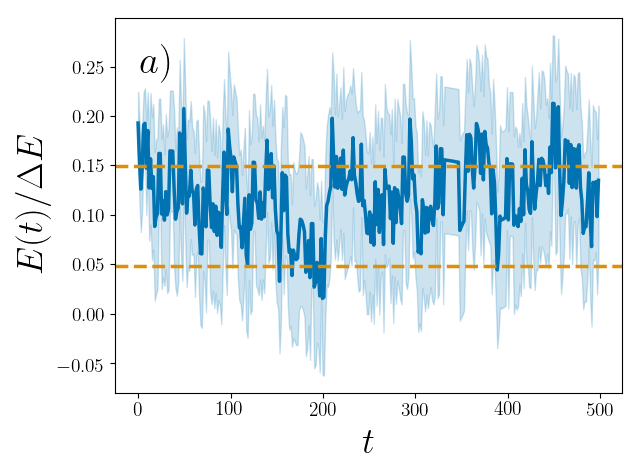}
	\includegraphics[width=0.32\textwidth]{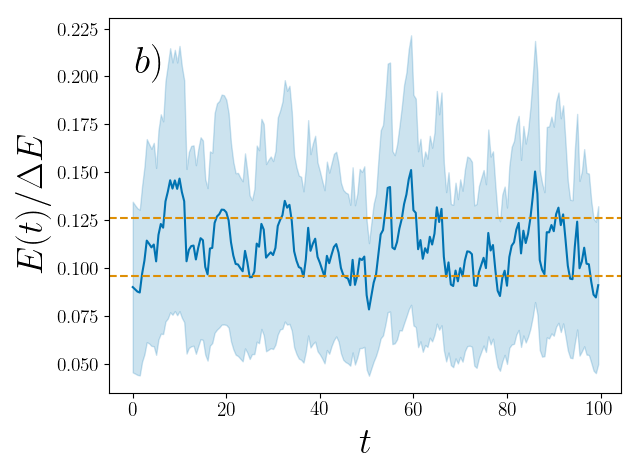}
	\includegraphics[width=0.32\textwidth]{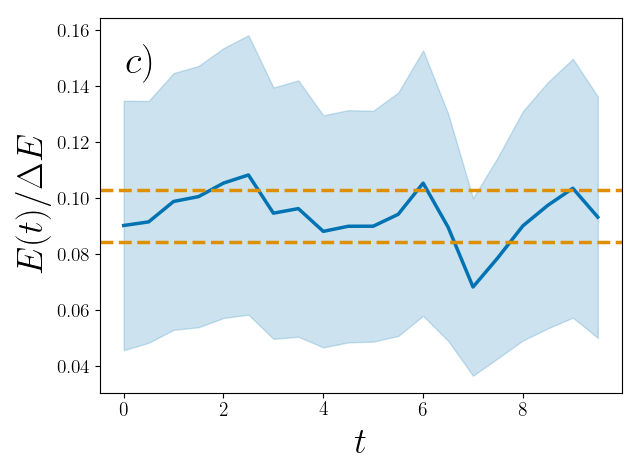}
	\caption{Change of total energy $E(t) = \langle H(t) \rangle$ in time due to action of repeated projective measurements (blue solid line). Variance of energy $\sigma_E(t)$ shown in shaded area. Dashed lines show time-fluctuations of energy $\delta_E(\infty)$. $\Delta E = E_{max} - E_{min}$. a) Quantum harmonic oscillator chain of main text, Parameters: $J = 0.8, h_x = 0.7, S = 3, N = 4$. Time averages variance $\overline{\sigma_E(t)}$ of order $\frac{\overline{\sigma_E(t)}}{\Delta E} \approx 0.05$ b) and c) Spin chain of Eq. \eqref{eq:SC_Hamiltonian} under action of local observable $S_z^1$ and global observable $\sum_i S_z^i$ respectively. $\frac{\overline{\sigma_E(t)}}{\Delta E} \approx 0.015, 0.009$ for local and global observables respectively. Parameters: $N = 4, S = 3, h_z = 1, h_x  =0.2, J = 0.8, \Delta = 0.3, q = 1.5$ }
	\label{fig:energies} 
\end{figure*}

Here we give some additional numerical results in order to verify the results of the main text. First, we note that an assumption made above is that the energy does not change in time significantly due to the action of measurements in a quantum jump trajectory. This is a reasonable assumption in the limit of a very large bath, where the system contributes little to the total energy. We confirm this assumption for the numerical models studied, where the bath is of a modest size, in Fig. \ref{fig:energies}.

\subsection{Measurement of the Density of States}\label{App:MeasureDOS}

In the main text we obtained the fluctuation relation
\begin{equation}\label{eq:DOS_flucs_Supp}
\frac{\sigma_O^2(\infty)}{\delta_O^2(\infty)} = 4 \pi D(E) \Gamma,
\end{equation}
which we show in this section via numerical exact diagonalizations may be exploited to measure the density of states of a quantum system. We show this for two models, the first is the quantum harmonic oscillator model of the main text. The second model we use is a chain of spin-$\frac{1}{2}$ particles, which more closely resembles an ion chain or other system of qubits. Eq. \ref{eq:DOS_flucs_Supp} applies to such models, as this relation does not require any assumptions on the system observable other than the requirement that it is diagonal in the non-interacting eigenbasis, and thus a large system dimension is not required.

The spin-$\frac{1}{2}$ chain is described by a
Hamiltonian of the form,
\begin{equation}\label{eq:spin_chain_Hamil}
H = H_S + H_B + H_{SB},
\end{equation}
where $H_S$ describes a single spin in a $B_z$ field
\begin{equation}
H_S = B_{z}^{(S)}\sigma_z^{(1)}.
\end{equation}
Here $\{\sigma_i^{(j)}\}\quad i = {x, y, z}$ are the Pauli operators acting on site $j$. We take the system as site $j = 1$. The bath Hamiltonian is a spin-chain of length $N-1$, with nearest-neighbour Ising and XX interactions subjected to both $B_z$ and $B_x$ fields
\begin{equation}
\begin{split}
H_B &= \sum_{j > 1}^N( B_{z}^{(B)}\sigma_z^{(j)}  + B_{x}^{(B)}\sigma_x^{(j)}) \\& +\sum_{j > 1}^{N-1}(J_z\sigma_z^{(j)}\sigma_z^{(j+1)} + J_x(\sigma_+^{(j)}\sigma_-^{(j+1)} + \sigma_-^{(j)}\sigma_+^{(j+1)} )).
\end{split}
\end{equation}
The interaction Hamiltonian describes the coupling of the system spin to a single central bath ion of index $N_{\rm m} = 3$,
\begin{equation}
H_{SB} = 
J_z^{(SB)}\sigma_z^{(1)}\sigma_z^{(N_{\rm m})} + J_x^{(SB)}(\sigma_+^{(1)}\sigma_-^{(N_{\rm m})} + \sigma_-^{(1)}\sigma_+^{(N_{\rm m})}).
\end{equation}
For the initial state of the spin-$\frac{1}{2}$ system we choose a randomly selected eigenstate of $H_0 = H_S + H_B$, ensuring only that the initial system state is $|\uparrow\rangle$, and that the initial energy is in the central $\frac{1}{2}$ of the total energy spectrum (guaranteeing that it is not too close to the ground state).

\begin{figure*}
	\includegraphics[width=0.49\textwidth]{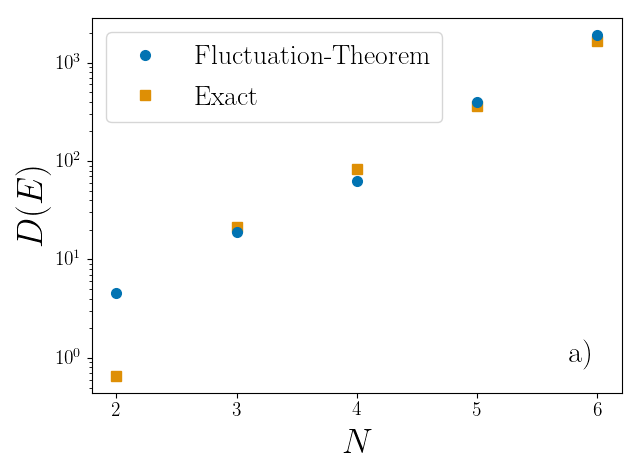}
	\includegraphics[width=0.49\textwidth]{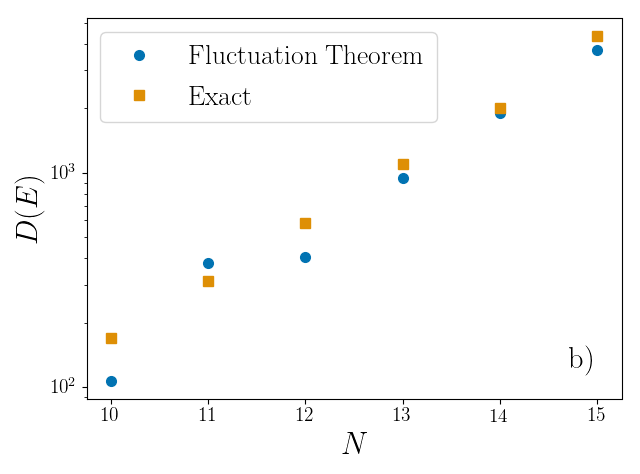}
	\caption{Comparison of density of states as inferred from the fluctuation theorem of Eq. \eqref{eq:DOS_flucs_Supp} (yellow squares) with exact value (blue circles). a) For system of coupled quantum harmonic oscillators described in main text. Parameters: $J = 1.2$, $h_x = 0.8$, $S=2$ b) For chain of spin-$\frac{1}{2}$ particles with Hamiltonian \eqref{eq:spin_chain_Hamil}. Parameters: $B^{(S)}_x = 0,    B^{(B)}_z = 0, B^{(B)}_x = 0.3, J^{(S)}_x = 0.4, J^{(S)}_z = 0.2, J^{(B)}_z = 0.1, J^{(B)}_x = 1, B^{(S)}_z = 0.8$. }
	\label{fig:measure_DOS} 
\end{figure*}

Numerics confirming Eq. \eqref{eq:DOS_flucs_Supp} are shown for both the quantum harmonic oscillator of the main text and the above spin-$\frac{1}{2}$ chain in Figs. \ref{fig:measure_DOS} a) and b) respectively.

In the exact diagonalization calculations in Fig. \ref{fig:measure_DOS} we have calculated $D(E)$ in two ways. The first is the exact value obtained numerically, and the second is a numerical experiment performed by calculating the  $\sigma_O^2$ and $\delta_O^2$ and $\Gamma$ from the expectation value dynamics $\langle O(t) \rangle$ of a local observable ($\sigma_z$ for the chain of spin-$\frac{1}{2}$ particles, the same as in the main text for the remaining models). Each of these three quantities are obtainable in a realistic experimental scenario, and thus this approach may be exploited in order to measure the DOS of a many-body quantum system.

\section{Ornstein-Uhlenbeck Process}\label{App:OU}

In the main text we make comparisons of the results to the classical dynamics of the Ornstein-Uhlenbeck (OU) process, describing the Brownian motion \cite{Einstein1905} of the position $x(t)$ of a particle in a medium subjected to random collisions with its environment. We summarize the relevant results here, and show a modification that reproduces the same finite-size time-fluctuations as the RMT model of the main text.

The OU process is described by the Langevin equation,
\begin{equation}\label{eq:OU_LV} 
\frac{d x(t)}{dt} = - \frac{k}{\gamma} (x(t) - \overline{x}) + \xi(t),
\end{equation}
where $\gamma$ and $\overline{x}$ are constants, and $\xi (t)$ is a stochastic random variable fulfilling 
$\langle \xi(t) \rangle_\xi = 0$, $\langle \xi(t) \xi(t^\prime)\rangle_\xi = 2D \delta(t - t^\prime)$, where $\langle \cdots \rangle_\xi$ indicates an average over stochastic trajectories, and $D$ is the diffusion constant.
The OU process describes the motion of an overdamped harmonic oscillator driven by white noise, with an oscillator potential $V(x) = \frac{k}{2}x^2$.
This is easily solved \cite{Uhlenbeck1930, Doob1942} to find (setting $\overline{x} = 0$) $x(t) = x(0)e^{-\frac{k}{\gamma} t}$ and $\langle x^2(t) \rangle_\xi = \langle x(t)\rangle_\xi^2 + \frac{D\gamma}{k}(1 - e^{-2\frac{k}{\gamma} t})$. The long-time observable variance may be written as
$\sigma_x^2(\infty) = \frac{D \gamma}{k}$.
For a system in thermal equilibrium, the time-average energy is
$\overline{\langle E \rangle} = \frac{1}{2}k_B T$ 
by the equipartition theorem. 
We then see that the long-time average energy gives $\overline{\langle V \rangle} = \frac{1}{2}D \gamma$, such that $D = \frac{k_B T}{\gamma}$. This is the celebrated Einstein relation of Brownian motion, a manifestation of the fluctuation-dissipation theorem (FDT) \cite{Kubo1966}. Note that in the case of the OU process the equipartition theorem is invoked in order to obtain the Einstein relation, whereas for our description in terms of chaotic wavefunctions both can be observed to emerge simultaneously, and are encompassed in Eq. (11) of the main text.

We further note that if one modifies the stochastic noise $\xi(t)$ such that $\langle \xi(t) \rangle_\xi = v(s)$, with $v$ a random variable itself, with $\overline{v(s)} = 0$, and $\overline{v(s)v(s^\prime)} = v^2 \delta(s - s^\prime)$, we obtain,
\begin{equation}
\delta_x^2(\infty) = \frac{v^2\gamma^2}{k^2}.
\end{equation}
Note that the physical interpretation of this modification is a shaking of the harmonic trap with white noise at a random velocity $v$ for any given realization of the random force $\xi$. In this case, we can make the association
\begin{equation}
v^2 \Rightarrow  \frac{\overline{[\Delta O^2]}_{\alpha_0}\Gamma}{4\pi D(E) } = \frac{k_B T\Gamma}{4\pi D(E) m}.
\end{equation}
The modified time-fluctuation can be thought of as an equivalent of the Einstein relation for time-fluctuations of finite classical systems.

%\bibliographystyle{apsrev4-1}
%\bibliography{bibli}

\end{document}